\documentclass[11pt]{article}

\usepackage{amsmath}
\usepackage{amssymb}
\usepackage{graphicx}
\usepackage{booktabs}
\usepackage[numbers]{natbib}
\usepackage{hyperref}
\usepackage[margin=1in]{geometry}

\title{HTM-EAR: Importance-Preserving Tiered Memory with Hybrid Routing under Saturation}
\author{Shubham Kumar Singh \\ February 2026}
\date{}

\begin{document}

\maketitle

\begin{abstract}
Memory constraints in long-running agents require efficient management of facts while preserving essential information. We present HTM-EAR, a tiered memory substrate that combines HNSW-based working memory (L1) and archival storage (L2) with importance-aware eviction and hybrid routing. When L1 reaches capacity, items are evicted using a weighted score of importance and usage. Queries first search L1; if similarity or entity coverage is insufficient, they fall back to L2. Retrieved candidates are re-ranked by a cross-encoder. We evaluate the system under saturation (15,000 facts, L1 capacity 500, L2 capacity 5000) using synthetic data across five seeds and on real BGL logs. Ablation studies compare full system against variants without cross-encoder, without gating, with LRU eviction, and an oracle with unbounded memory. Results show that the full model preserves active query precision (MRR = 1.000) while forgetting older history, approaching oracle active performance (0.997 ± 0.003). LRU minimizes latency (21.1 ms) but loses essential information (2416 essential facts evicted). On BGL logs, full system achieves MRR 0.336, close to oracle 0.370, while LRU drops to 0.069. The code is available at \url{https://github.com/shubham-61291/HTM-EAR}.
\end{abstract}

\section{Introduction}

Autonomous agents operating over extended periods accumulate vast amounts of information, yet computational memory is bounded. A common solution is to maintain a small working memory and archive older items, but naive eviction (e.g., LRU) can discard critical facts.

Recent work in approximate nearest neighbor retrieval using HNSW~\cite{hnsw} and embedding-based ranking via Sentence-BERT~\cite{sentence-transformers} has enabled efficient semantic search. ANN benchmarks~\cite{ann-survey} have shown that HNSW offers competitive efficiency. However, these approaches typically assume unlimited memory or static corpora. The challenge of maintaining a bounded working memory while preserving essential information under streaming conditions remains under-explored. Our work addresses this gap by introducing a tiered memory architecture with importance-aware eviction and hybrid routing.

This paper introduces HTM-EAR, a system that separates memory into two tiers: a fast, limited L1 and a larger L2. Eviction from L1 is guided by an importance score combined with access frequency. Queries are routed adaptively: if L1 cannot provide a sufficiently similar result or fails to cover query entities, the search falls back to L2. Retrieved items are finally re-ranked by a cross-encoder for high precision. We evaluate the system under extreme saturation and compare several ablations to understand the contribution of each component.

\section{System Architecture}

\begin{figure}[h]
\centering
\includegraphics[width=0.9\linewidth]{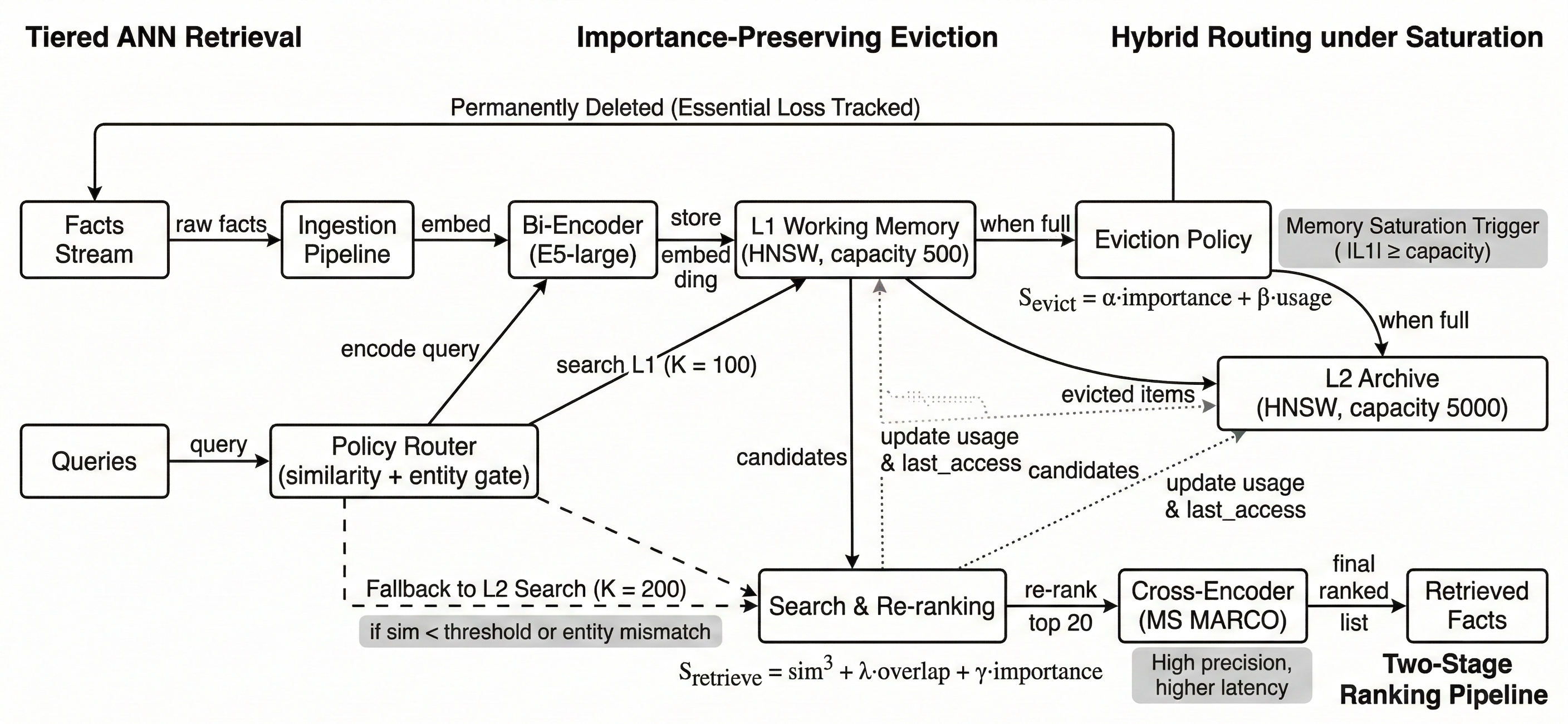}
\caption{HTM-EAR architecture illustrating tiered ANN retrieval, importance-aware eviction, and hybrid routing under memory saturation.}
\label{fig:architecture}
\end{figure}

Figure~\ref{fig:architecture} illustrates the HTM-EAR architecture. Incoming facts are embedded by a bi-encoder (E5-large)~\cite{sentence-transformers} and stored in L1 working memory, an HNSW index~\cite{hnsw} with capacity 500. When L1 becomes full, an eviction policy selects items to move to L2, an HNSW archive of capacity 5000. If L2 also reaches capacity, items are permanently deleted and tracked as essential loss if their importance exceeds a threshold (0.85).

Queries are encoded and first search L1 (k=100). A policy router compares the top result's similarity to a threshold (0.84) and checks whether all query entities appear in the item's entity set. If either condition fails, the query also searches L2 (k=200). Retrieved candidates are scored by a combination of similarity, entity overlap, and importance, then the top 20 are re-ranked using a cross-encoder trained on MS MARCO~\cite{msmarco} to produce the final list.

\section{Retrieval and Eviction Formulation}

\textbf{Eviction scoring.} When L1 reaches capacity, items are scored for eviction as
\[
S_{\text{evict}} = \alpha \cdot \text{importance} + \beta \cdot \min\left(\frac{\text{usage}}{10}, 1\right),
\]
with $\alpha = 0.75$, $\beta = 0.25$. Items with the lowest scores are evicted (approximately 15\% of capacity). Evicted items are inserted into L2, which may itself evict using the same policy.

\textbf{Retrieval scoring.} During retrieval, candidate items from L1 or L2 are assigned a score
\[
S_{\text{retrieve}} = \text{sim}^3 + \lambda \cdot \text{overlap} + \gamma \cdot \text{importance},
\]
where $\text{sim}$ is the inner product similarity from HNSW, $\text{overlap}$ is the number of common entities between query and item, and $\lambda = 0.8$, $\gamma = 0.1$. The cubic transformation emphasizes high-similarity matches while suppressing marginal ones.

\section{Experimental Setup}

We generate synthetic datasets of $N$ facts each with a corresponding query. Facts are assigned importance 0.95 if they contain certain keywords (e.g., ``panic''), otherwise 0.5. For the saturation scenario (Scenario B), $N = 15{,}000$, L1 capacity 500, L2 capacity 5000. We evaluate five system modes:

\begin{itemize}
\item \texttt{full}: the complete system as described.
\item \texttt{oracle\_unbounded}: a single unbounded memory (no eviction).
\item \texttt{no\_ce}: full system without cross-encoder re-ranking.
\item \texttt{no\_gate}: full system without the similarity/entity gating; only L1 is searched.
\item \texttt{lru}: both tiers use LRU eviction (based on last access time) instead of importance-aware scoring.
\end{itemize}

For each mode we run five seeds (42–46). Evaluation measures Mean Reciprocal Rank (MRR) separately on the most recent 100 facts (``Active'') and the first 100 facts (``History''). We also record latency per query and the number of essential facts lost (importance $\geq$ 0.85 that are permanently deleted). Additionally, we evaluate on real BGL logs~\cite{bgl} (2,000 entries) with queries generated from extracted entities.

\section{Ablation Study under Saturation (Scenario B)}

Table~\ref{tab:retrieval} shows MRR for each mode on active and history facts. The full model achieves perfect active MRR (1.000) but low history MRR (0.215), indicating that important recent facts are retained while older ones are forgotten. In contrast, the oracle unbounded retains both active and historical facts (0.997 active, 0.990 history). Removing the cross-encoder (\texttt{no\_ce}) hardly changes performance. Removing the gate (\texttt{no\_gate}) degrades active MRR to 0.432, confirming that fallback to L2 is essential for saturation. LRU also gives perfect active MRR but completely loses history (0.000) because it does not differentiate importance.

\begin{table}[h]
\centering
\begin{tabular}{@{}lcc@{}}
\toprule
Mode & Active MRR (mean$\pm$std) & History MRR (mean$\pm$std) \\
\midrule
full          & $1.000 \pm 0.000$ & $0.215 \pm 0.028$ \\
lru           & $1.000 \pm 0.000$ & $0.000 \pm 0.000$ \\
no\_ce        & $1.000 \pm 0.000$ & $0.218 \pm 0.029$ \\
no\_gate      & $0.432 \pm 0.025$ & $0.000 \pm 0.000$ \\
oracle\_unbounded & $0.997 \pm 0.003$ & $0.990 \pm 0.005$ \\
\bottomrule
\end{tabular}
\caption{Retrieval performance (MRR) under saturation. Active refers to the last 100 facts, History to the first 100.}
\label{tab:retrieval}
\end{table}

\begin{figure}[h]
\centering
\includegraphics[width=0.9\linewidth]{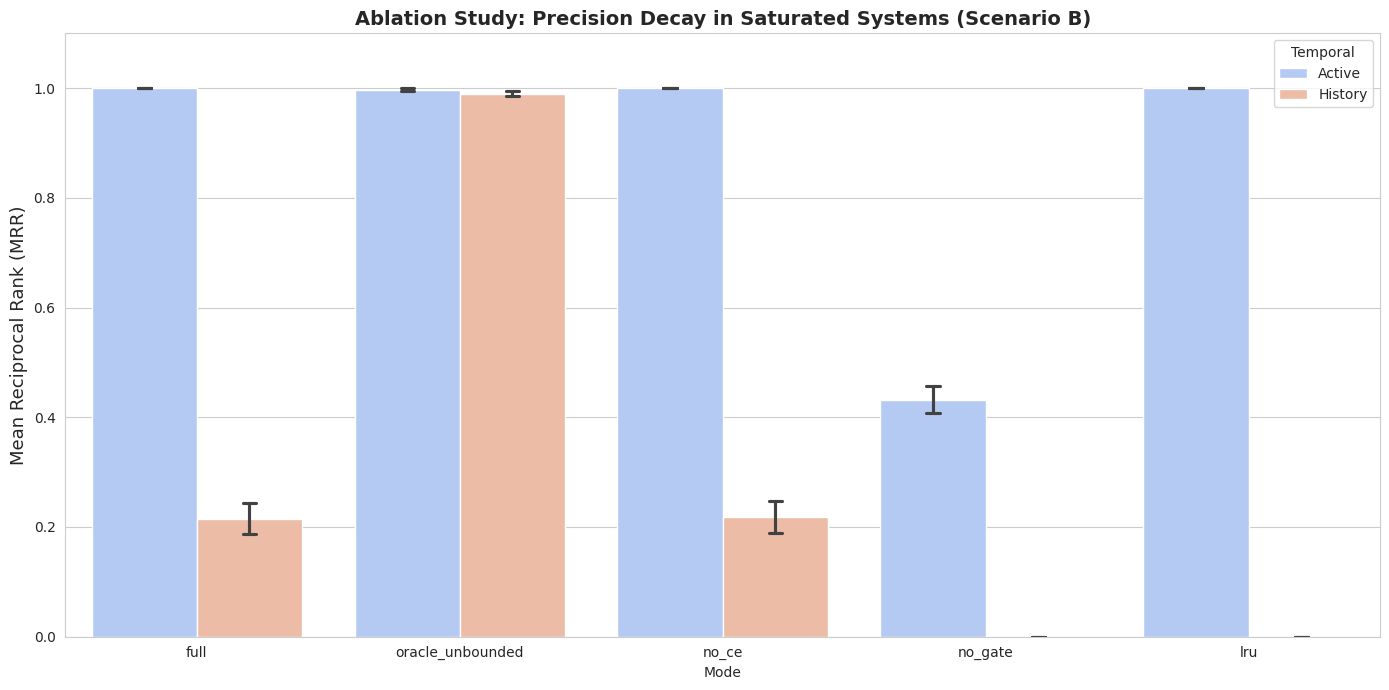}
\caption{Precision decay under saturation (Scenario B).}
\label{fig:ablation}
\end{figure}

\section{Pareto Tradeoff Analysis}

Table~\ref{tab:robustness} reports latency, essential loss, and total pruned items for active queries. The full model has latency 39.7 ms and loses no essential facts. LRU is fastest (21.1 ms) but loses 2416 essential facts on average. \texttt{no\_ce} achieves similar low latency without essential loss. \texttt{no\_gate} has the highest latency (41.1 ms), reflecting inefficient retrieval under saturation. The oracle unbounded has moderate latency (37.4 ms) and no eviction.

\begin{figure}[h]
\centering
\includegraphics[width=0.9\linewidth]{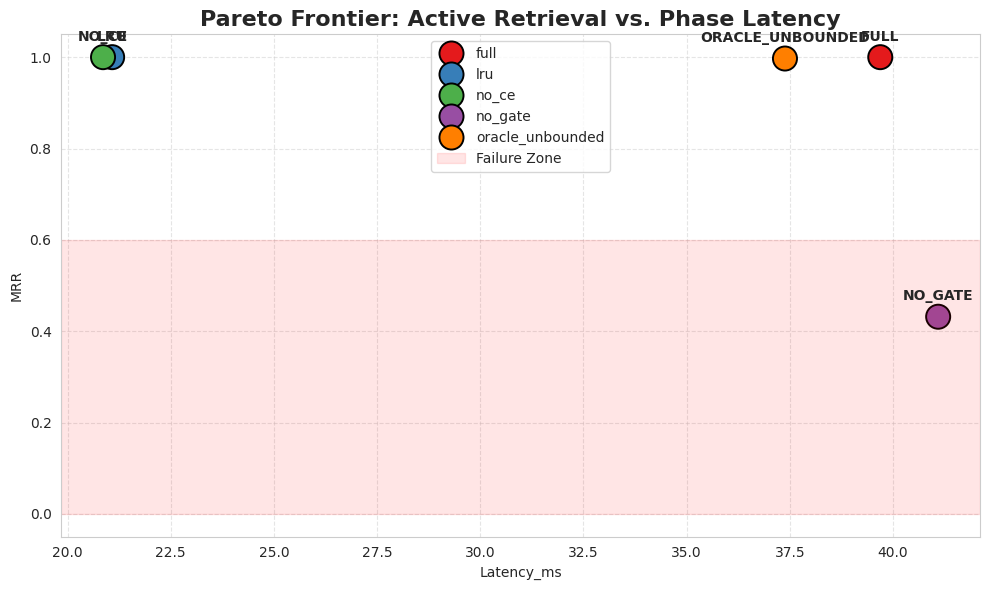}
\caption{Pareto frontier between active retrieval quality and latency. The failure zone indicates MRR below 0.6.}
\label{fig:pareto}
\end{figure}

Figure~\ref{fig:pareto} plots the tradeoff between latency and active MRR. The full model lies near the oracle, while LRU sacrifices essential information for speed. \texttt{no\_ce} provides a good balance, and \texttt{no\_gate} falls into a failure zone due to low MRR.

\begin{table}[h]
\centering
\begin{tabular}{@{}lccc@{}}
\toprule
Mode & Latency (ms) & Essential Lost & Pruned Total \\
\midrule
full          & $39.69 \pm 3.08$ & $0.0 \pm 0.0$   & $9750 \pm 0$ \\
lru           & $21.08 \pm 3.26$ & $2416.4 \pm 23.1$ & $9750 \pm 0$ \\
no\_ce        & $20.86 \pm 1.69$ & $0.0 \pm 0.0$   & $9750 \pm 0$ \\
no\_gate      & $41.09 \pm 1.56$ & $0.0 \pm 0.0$   & $9750 \pm 0$ \\
oracle\_unbounded & $37.38 \pm 2.90$ & $0.0 \pm 0.0$   & $0 \pm 0$ \\
\bottomrule
\end{tabular}
\caption{System robustness metrics (mean $\pm$ std) for active queries.}
\label{tab:robustness}
\end{table}

\section{Real-World Log Validation (BGL)}

\begin{figure}[h]
\centering
\includegraphics[width=0.9\linewidth]{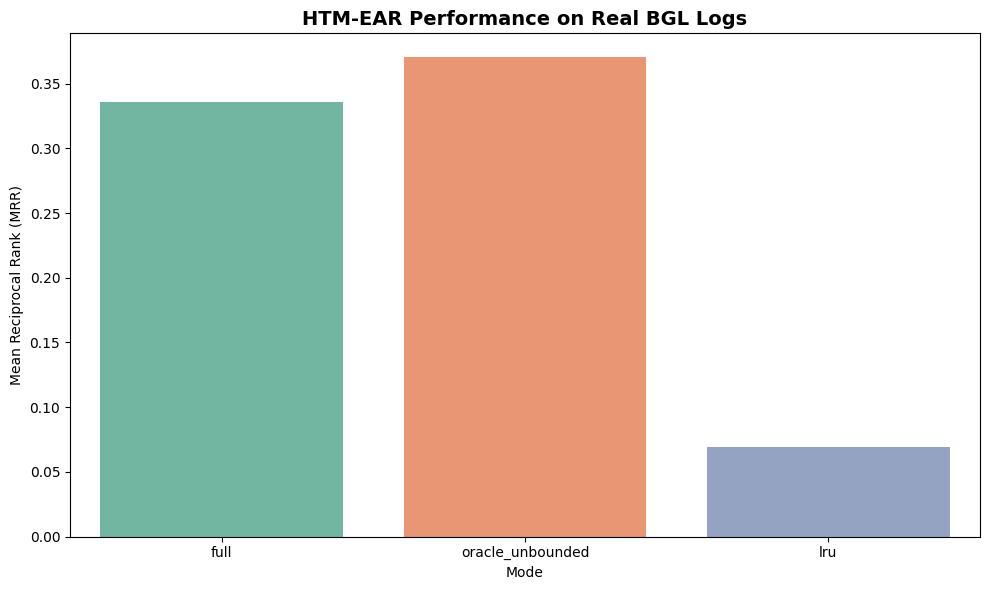}
\caption{Performance comparison on BGL log benchmark.}
\label{fig:bgl}
\end{figure}

We evaluate the three most representative modes on 2,000 entries from the BGL log dataset~\cite{bgl}. Results are shown in Table~\ref{tab:bgl} and Figure~\ref{fig:bgl}. The full system achieves MRR 0.336, close to the oracle unbounded (0.370). LRU performs poorly (0.069), confirming that importance-agnostic eviction harms retrieval on realistic logs where query entities are sparsely distributed.

\begin{table}[h]
\centering
\begin{tabular}{@{}lccc@{}}
\toprule
Mode & MRR & Latency (ms) & Essential Lost \\
\midrule
full          & 0.336 & 42.70 & 0 \\
oracle\_unbounded & 0.370 & 42.04 & 0 \\
lru           & 0.069 & 22.49 & 0 \\
\bottomrule
\end{tabular}
\caption{Performance on BGL log benchmark.}
\label{tab:bgl}
\end{table}

\section{Discussion}

The experiments reveal several insights. First, importance-aware eviction (\texttt{full}) preserves essential information while LRU discards it, even though both achieve perfect active MRR. Second, the hybrid router (similarity + entity gate) is critical: without it, many queries fail to find relevant facts in L1 and never consult L2, causing a sharp MRR drop. Third, the cross-encoder adds latency but does not significantly improve MRR under this synthetic setup; however, it may be more valuable on complex real-world queries. Fourth, the oracle unbounded demonstrates an upper bound that the full system approaches on active queries, indicating that the tiered design is effective under saturation.

Latency varies across modes: LRU and \texttt{no\_ce} are fastest because they avoid the cross-encoder or L2 fallback. The full model and \texttt{no\_gate} have similar latency, but \texttt{no\_gate} suffers from poor recall. The Pareto analysis shows that \texttt{no\_ce} offers the best latency–precision tradeoff among bounded-memory systems.

\section{Limitations}

This study has several limitations. The synthetic data, while controlled, may not capture the full complexity of real-world fact distributions. Capacities were fixed at 500 and 5000; performance under different sizes remains unexplored. Eviction and retrieval weights ($\alpha,\beta,\lambda,\gamma$) were chosen heuristically and not tuned. Real-world evaluation is limited to a single log dataset (BGL); broader benchmarks are needed. The system does not adapt thresholds online, and no theoretical guarantees are provided.

\section{Reproducibility and Code Availability}

The complete implementation is publicly available at \url{https://github.com/shubham-61291/HTM-EAR}. The file \texttt{htm\_ear.py} contains all code for dataset generation, system implementation, and evaluation. All experiments can be reproduced by running this script; results in this paper were obtained with Python 3.9 and the dependencies listed in the code.

\section{Conclusion}

We presented HTM-EAR, a tiered memory system with importance-aware eviction and hybrid routing for long-running agents. Under saturation, the system preserves essential information and approaches oracle active performance on recent facts while gracefully forgetting older ones. Ablation studies isolate the contributions of each component: importance scoring prevents essential loss, hybrid routing ensures recall, and the cross-encoder provides marginal gain in this setting. Real-world validation on BGL logs confirms the approach's viability. Future work may explore adaptive parameter tuning and larger-scale production deployments.

\end{document}